\documentclass[atmp]{ipart_v1}
\Year{2020}
\usepackage{t1enc}
\usepackage[english]{babel}

\usepackage{amsthm}
\usepackage{yfonts}

\usepackage{bbm}
\usepackage{bm}
\usepackage{mathrsfs}

\usepackage{amsmath,amssymb,amsfonts,bm,amscd}
\usepackage{graphicx}
\usepackage{subcaption}
\usepackage{centernot}
\usepackage{verbatim}
\usepackage{fancybox}
\usepackage{slashed}
\usepackage{url}
\usepackage{array}
\usepackage{hhline}
\usepackage{tabularx}
\usepackage{tikz}
\usepackage[utf8]{inputenc}

\graphicspath {{figures/}}

\newcommand{\bea}{\begin{eqnarray}}
\newcommand{\eea}{\end{eqnarray}}

\newcommand{\Z}{{\mathbb Z}}

\newcommand{\C}{{\mathbb C}}

\newcommand{\fq}{{\frak q}}

\newcommand{\rnum}[1]{\lowercase\expandafter{\romannumeral #1\relax}}
\newcommand{\Rnum}[1]{\uppercase\expandafter{\romannumeral #1\relax}}
\def\Tr{{\rm Tr \,}}

\def\half{\frac{1}{2}}

\def\ft{\frak{t}} %equivariant parameter
\def\fu{\frak{u}} %equivariant parameter

\def\rq{q}
\def\rt{t}

\def\frak{\mathfrak}

\def\tilde{\widetilde}

\def\CF{{\mathcal F}}

\def\CH{{\mathcal H}}
\def\CI{{\mathcal I}}

\def\CM{{\mathcal M}}
\def\CN{{\mathcal N}}

\def\CS{{\mathcal S}}
\def\CT{{\mathcal T}}

\makeatletter
\DeclareRobustCommand*\cal{\@fontswitch\relax\mathcal}
\makeatother

\def\^{{\wedge}}
\def\*{{\star}}

\newcommand{\beq}{\begin{equation}\begin{aligned}}
\newcommand{\eeq}{\end{aligned}\end{equation}}

\newcommand{\be}[0]{\begin{equation}}
\newcommand{\ee}[0]{\end{equation}}

\numberwithin{equation}{section}

\theoremstyle{plain}% default

\begin{document}

\title[A.-D. Theories, Modularity of Minimal Models and Refined C.-S.]{Argyres-Douglas Theories, Modularity of Minimal Models and Refined Chern-Simons}

\author[Can Koz\c{c}az, Shamil Shakirov, Wenbin Yan]{Can Koz\c{c}az, Shamil Shakirov, Wenbin Yan}

\begin{abstract}
The Coulomb branch indices of Argyres-Douglas theories on $L(k,1)\times S^{1}$ are recently identified with matrix elements of modular transforms of certain $2d$ vertex operator algebras in a particular limit.  A one parameter generalization of the modular transformation matrices of $(2N+3,2)$ minimal models are proposed to compute the full Coulomb branch index of $(A_{1},A_{2N})$ Argyres-Douglas theories on the same space. Moreover, M-theory construction of these theories suggests direct connection to the refined Chern-Simons theory. The connection is made precise by showing how the modular transformation matrices of refined Chern-Simons theory are related to the proposed generalized ones for minimal models and the identification of Coulomb branch indices with the partition function of the refined Chern-Simons theory.
\end{abstract}

\maketitle

\section{Introduction}
\label{sec:intro}

Generalized Argyres-Douglas (AD) theories and their construction from M5 branes \cite{Argyres:1995jj, Gaiotto:2009hg, Bonelli:2011aa, Xie:2012hs, Wang:2015mra} lead to various predictions in mathematics. On the one hand, their Coulomb branch moduli spaces are identified with moduli spaces of wild Hitchin systems \cite{Gaiotto:2009hg, Xie:2012hs}. On the other hand, the correspondence between $4d$ $\CN=2$ superconformal field theories (SCFTs) and $2d$ vertex operator algebras (VOAs) \cite{Beem:2013sza} can also be applied to AD theories, relating them with minimal models, Kac-Moody algebras and other VOAs \cite{Cordova:2015nma, Song:2015wta, Buican:2015ina, Buican:2016arp, Xie:2016evu, Creutzig:2017qyf,Song:2017oew}. Hence, it is possible to use AD theories as a bridge to study the possible connections between wild Hitchin systems and VOAs \cite{Gukov:2016lki,Fredrickson:2017yka}.

\subsection*{AD theories, wild Hitchin systems and VOAs}

An AD theory $\CT$ can be constructed by compatifying $6d$ $(2,0)$ SCFT of type $G=ADE$ on a sphere $\Sigma$ with one irregular singularities and possible regular singularities \cite{Gaiotto:2009hg, Bonelli:2011aa, Xie:2012hs, Wang:2015mra}. The Coulomb branch $\CM_{\CT}$ of $\CT$ compactifed on $S^1$ is the Hitchin moduli space $\CM_H(\Sigma,G)$ \cite{Gaiotto:2009hg, Xie:2012hs}, whose mirror ${^L}\!{\CM}_{\CT}$ is given by $\CM_H(\Sigma,{^L}\!G)$ associated with the Langlands dual group ${^L}\!{G}$ via the geometric Langlands correspondence \cite{beilinson1991quantization,hausel2003mirror,Kapustin:2006pk,Gukov:2006jk}. This was verified by matching the lens space Coulomb index of AD theories and the wild Hitchin characters \cite{Gukov:2016lki,Fredrickson:2017yka},
\be
{\cal I}_{\rm Coulomb} (\CT[\Sigma, G]; L(k,1)\times S^1)= \dim_{\ft} \CH(\Sigma, {^L}G_{\mathbb{C}}; k),
\label{EVF=CBI}
\ee
where $\CH(\Sigma, {^L}G_{\mathbb{C}}; k)$ is the Hilbert space of complex Chern-Simons (CS) theory that is obtained by quantizing the Hitchin moduli space.

In \cite{Fredrickson:2017yka, Fredrickson:2017jcf}, $2d$ VOAs are added to the previous relations to make it into a triangle,
\be\label{Triangle}
\begin{array}{rcl}
\text{Coulomb index of $\CT$} & \longleftrightarrow & \text{quantization of ${^L}{\CM}_{\CT}$}
\\
\\
$\rotatebox[origin=c]{-45}{$\longleftrightarrow$}$   &  & $\rotatebox[origin=c]{45}{$\longleftrightarrow$} $ \\
\\
& \text{VOA $\chi_{\CT}$}
&
\end{array}
\ee
where the VOA $\chi_\CT$ associated with the 4d $\CN=2$ theory $\CT$. It is observed that the fixed points of $U(1)$ Hitchin action on $\CM_\CT$ are in bijection with highest-weight representations of $\chi_\CT$. In addition, a particular limit of the Coulomb index (or the Hitchin character) can be expressed in terms of modular transformation matrices of those representations. The striking feature here is that the VOA $\chi_\CT$ is usually related to Schur operators and Higgs branch of $\CT$ \cite{Beem:2013sza, Cordova:2015nma, Song:2015wta, Buican:2015ina, Buican:2016arp, Xie:2016evu, Creutzig:2017qyf,Song:2017oew,Fluder:2017oxm}, which do not contain Coulomb branch at all!

However, the relation between the Coulomb branch index (wild Hitchin characters) of $\CT$ and modular transformation matrices of $\chi_\CT$ in \cite{Fredrickson:2017yka} is not yet complete. Because the Coulomb branch index depends on a fugacity $\fu$ which counts the $U(1)_r$ charge of the 4d ${\cal N}=2$ superconformal algebra, while elements the modular transformation matrices of $\chi_\CT$ are numbers. The relation holds only when $\fu$ approaches a special value given below.

It is this current work's goal to construct the full relation between Coulomb branch indices of $(A_1, A_{2N})$ AD theories and modular transformation matrices of minimal models. We conjecture that Coulomb branch indices of $(A_1, A_{2N})$ AD theories on lens space $L(k,1)$ times a circle can be written as (up to a proportional constant),
\begin{equation}
\label{eq:introA1A2N}
\CI_{(A_1, A_{2N})}(\fu) \propto\left (\CS_{(A_1, A_{2N})}(\fu)^{-1}\CT_{(A_1, A_{2N})}^{-k}(\fu)\CS_{(A_1, A_{2N})}(\fu)\right)_{00}.
\end{equation}
where $\CS_{(A_1, A_{2N})}(\fu)$ and $\CT_{(A_1, A_{2N})}(\fu)$ are matrices with one parameter $\fu$ which satisfies the following relations,
\begin{equation}
\begin{split}
&\CS_{(A_1, A_{2N})}(\fu)^2=1,\\
&\left(\CS_{(A_1, A_{2N})}(\fu)\CT_{(A_1, A_{2N})}(\fu)\right)^3=1.
\end{split}
\end{equation}
Clearly $\CS_{(A_1, A_{2N})}(\fu)$ and $\CT_{(A_1, A_{2N})}(\fu)$ form a representation of $SL(2,\mathbb{Z})$, and the relation in \cite{Fredrickson:2017yka}\footnote{In fact, a slightly modified relation from \cite{Fredrickson:2017yka} is used in this paper, see equation \ref{eq:CBIandSTrelationMod}.} can be recovered by taking the  limit $\fu\rightarrow\exp\left(-\frac{2i\pi}{2N+3}\right)$ under which $\CS_{(A_1, A_{2N})}(\fu)$ and $\CT_{(A_1, A_{2N})}(\fu)$ become the modular transformation matrices $S_{(2N+3,2)}$ and $T_{(2N+3,2)}$ of characters of $(2N+3,2)$ minimal models, respectively. $\CS_{(A_1, A_{2N})}(\fu)$ and $\CT_{(A_1, A_{2N})}(\fu)$ can be viewed as one parameter generalization of $S_{(2N+3,2)}$ and $T_{(2N+3,2)}$, and it will be shown in section \ref{sec:inter} that $\CS_{(A_1, A_{2N})}(e^{-\frac{2\pi i}{2M+3}})$ and $\CT_{(A_1, A_{2N})}(e^{-\frac{2\pi i}{2M+3}})$ are  modular transformation matrices of {\bf torus one-point conformal blocks} of $(2M+3,2)$ models. In short, the Coulomb branch index of the $(A_1, A_{2N})$ AD theory is related not only to the modular property of $(2N+3,2)$ minimal model but all the $(2M+3,2)$ minimal models with $M\geq N$.

\subsection*{AD theories and refined Chern-Simons theory}

The fact that Coulomb branch indices of $(A_1, A_{2N})$ theories can be written as $SL(2,\mathbb{Z})$ elements $S^{-1}T^{-k}S$ implies that these indices are related to the topological invariants of 3-manifolds, in particular the topological invariants of the lens space $L(k,1)$. One construction of $L(k,1)$ is gluing boundaries of two solid tori up to an $SL(2,\mathbb{Z})$ transformation which maps the $(1,0)$-cycle to the $(1,k)$-cycle of the other one. The correct $SL(2,\mathbb{Z})$ transformation is just $S^{-1}T^{-k}S$ up to framing factors $T^{n_{L,\,R}}$ which may be added to the left or right. This is exactly the structure of the Coulomb branch index in Eq. \ref{eq:introA1A2N}!

It is then interesting to see if the Coulomb branch index on $L(k,1)\times S^1$ as the partition function on $L(k,1)$ of a three dimensional topological theory. To find this topological theory, it is useful to go back to the M-theory construction. The $(A_1, A_{2N})$ AD theories are engineered by compactification of M5 branes on a sphere with one irregular singularity, which is equivalent to a disk with special boundary condition \cite{Gaiotto:2009hg, Bonelli:2011aa, Xie:2012hs, Wang:2015mra}. Topologically this is the same as wrapping M5 branes on $L(k,1)\times S^1$ times a cigar geometry, which is the same construction of {\bf the refined Chern-Simons theory} (refined CS) \cite{Aganagic:2011sg} in M-theory, based on earlier work \cite{Dijkgraaf:2006um}! It is then natural to identify Coulomb branch indices of $(A_1, A_{2N})$ AD theories on $L(k,1)\times S^1$ with the refined CS partition function on $L(k,1)$,
\begin{equation}
\CI_{(A_1,A_{2N})}(\fu)
=\fu^{\half N(N+1)k}
Z^{rCS}(L(k,1);\rq=\fu^{-2},\rt=\fu^{2N+1}).
\end{equation}
The behavior of the irregular singularity of $(A_1, A_{2N})$ dictates the relation between $\fu$ of Coulomb branch index and refined CS equivariant parameters $\rq$ and $\rt$. One can then use this relation to conjecture expressions of other observables of AD theories from the refined CS theory. Therefore a forth player is added to the previous triangular relation,
\be\label{quad}
\begin{array}{ccc}
\text{Coulomb index of $\CT$} & \longleftrightarrow & \text{quantization of ${^L}{\CM}_{\CT}$}
\\
\\
${\rotatebox[origin=c]{-90}{$\longleftrightarrow$}}$   & {\color{red}\rotatebox[origin=c]{-45}{$\longleftrightarrow$}} \hspace{-0.55cm}\rotatebox[origin=c]{45}{$\longleftrightarrow$} & ${\color{blue}\rotatebox[origin=c]{90}{$\longleftrightarrow$} ? } $ \\
\\
 \text{VOA $\chi_{\CT}$} & {\color{red}\longleftrightarrow }
& \text{refined CS partition function}
\end{array}
\ee

This paper is organized as follows: Section \ref{sec:background}, summarizes the background knowledge used in this work. The relation between Coulomb branch indices of $(A_1, A_{2N})$ AD theories and modular properties of torus one-point conformal block of minimal models are studied in section \ref{sec:inter}. In section \ref{sec:rCS}, both a physical argument and explicit computations are presented in order to show the identification of Coulomb branch indices of $(A_1, A_{2N})$ AD theories on $L(k,1)\times S^1$ and refined CS partition function on $L(k,1)$. Section \ref{sec:generalizations}, generalizes the relation found in section \ref{sec:rCS} and predicts other partition functions of $(A_1, A_{2N})$ theories using refined CS theory.

\section{Background information}
\label{sec:background}

\subsection{Coulomb branch index of $(A_1, A_{2N})$ AD theories}

The Coulomb branch index on $L(k,1)\times S^1$ is defined in terms of the trace over the Hilbert space on $L(k,1)$ \cite{Gadde:2011uv, Benini:2011nc, Razamat:2013jxa,Razamat:2013opa,Nieri:2015yia},
\begin{equation}
\label{eq:def:Coulomb}
\CI^{C}=\Tr_C (-1)^F \ft^{r-R},
\end{equation}
where the trace is taken over BPS states annihilated by both $\tilde{Q}_{1\dot{-}}$ and $\tilde{Q}_{2\dot{+}}$ of the $4d$ superconformal algebra. $F$ is fermionic number of the state, $R$ and $r$ are the $SU(2)_R$ and $U(1)_r$ charges of the $4d$ superconformal algebra, respectively. Note that $L(k,1)$ is a quotient of $S^3$ by $\mathbb{Z}_k\subset U(1)_{\mathrm{Hopf}}\subset SU(2)_{L}\subset SO(4)$, and both $\tilde{Q}_{1\dot{-}}$ and $\tilde{Q}_{2\dot{+}}$ transform trivially under $SU(2)_{L}$, the trace formula in Eq. \ref{eq:def:Coulomb} is well defined.

The Coulomb branch indices for $(A_1, A_{2N})$ AD theories on $L(k,1)\times S^1$ was first discussed in \cite{Fredrickson:2017yka} using the ``Lagrangian" proposed by \cite{Maruyoshi:2016tqk,Maruyoshi:2016aim,Agarwal:2016pjo, Agarwal:2017roi}. Here we simply quote the result,
\begin{equation}
\begin{split}
& \CI_{(A_1,A_{2N})} = \\
& \sum_{i=0}^N
\frac{\fu^{i(i+1)k/2}}
{\prod_{l=1}^i\left(1-\fu^{2(N+l+1)}\right)\left(1-\fu^{-2l+1}\right)
\prod_{l=i+1}^N\left(1-\fu^{2l+1}\right)\left(1-\fu^{2(N-l+1)}\right)
},
\end{split}
\end{equation}
where we replace the equivariant parameter $\ft$ in \cite{Fredrickson:2017yka} by $\fu=\ft^{\frac{1}{2N+3}}$ for later convenience.

The $(A_1, A_{2N})$ AD theories are closely related to the $(2N+3,2)$ minimal models. It was also shown in \cite{Fredrickson:2017yka} that the Coulomb index of the 4$d$ theories are related to the modular property of the characters of minimal models,
\begin{equation}
\label{eq:CBIandSTrelation}
\lim_{\fu\rightarrow e^{\frac{2\pi i}{2N+3}}}
\CI_{(A_1,A_{2N})}(\fu)
= e^{\left(\frac{1}{12}-\frac{1}{4(2N+3)}\right)\pi i k} \left(\CS \CT^k \CS\right)_{0,0}.
\end{equation}
$\CS$ and $\CT$ are the matrix representations of modular $S$ and $T$ transformations acting on the characters of $(2N+3,2)$ minimal model, and they are $(N+1)$ by $(N+1)$ matrices because of $N+1$ irreducible modules in $(2N+3,2)$ model. $\CS$ and $\CT$ can be expressed explicitly,
\begin{equation}
\label{eq:modularcha}
\begin{split}
&\CS_{r\rho}=\frac{2}{\sqrt{2N+3}}(-1)^{n+r+\rho}\sin\left(\frac{2\pi(r+1)(\rho+1)}{2N+3}\right), \\
&\CT_{r\rho}=\delta_{r\rho}e^{2\pi i(h_{r,\rho}-c/24)},
\end{split}
\end{equation}
where $r$ and $\rho$ run from $0$ to $N$ with $0$ understood as the vacuum module $(1,1)$. $c$ is the central charge of $(2N+3,2)$ model, and $h_{r,\rho}$ is the conformal weight, defined in Eq. \ref{confweight} below. One may also check the following relation is also true,
\begin{equation}
\label{eq:CBIandSTrelationMod}
\lim_{\fu\rightarrow e^{-\frac{2\pi i}{2N+3}}}
\CI_{(A_1,A_{2N})}(\fu)
= e^{\left(-\frac{1}{12}+\frac{1}{4(2N+3)}\right)\pi i k} \left(\CS \CT^{-k} \CS\right)_{0,0}.
\end{equation}
In this modified relation the limit $\fu$ is taken to be $e^{-\frac{2\pi i}{2N+3}}$ instead of $e^{\frac{2\pi i}{2N+3}}$ and $\CT^k$ is replaced by $\CT^{-k}$ for later conveniences.

This paper will discuss a more general relation between $\CI_{(A_1,A_{2N})}(\fu)$ and modular properties of minimal models. Knowledge beyond characters is required to achieve this.

\subsection{Torus one-point conformal blocks for $(p,q)$ minimal models}
\label{sec:torus1pt}

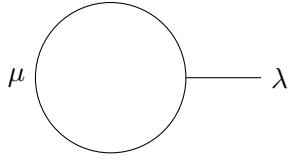
\begin{figure}
\centering
\begin{tikzpicture}
\draw (0,0) circle [radius=1];
\draw (1,0) -- (2,0);
\node [left] at (-1,0) {$\mu$};
\node [right] at (2,0) {$\lambda$};
\end{tikzpicture}
\caption{\label{fig:torus1pt}Schematics of the torus one-point conformal block $\CF^{\lambda}_{c,\mu}(e^{2\pi i \tau})$.}
\end{figure}

One natural generalization of characters are the torus one-point conformal \newline block $\CF^\lambda_{c,\mu}(e^{2\pi i \tau})$ as depicted in figure \ref{fig:torus1pt}, where $c$ is the central charge of the model, $\lambda$ is the conformal dimension of the external primary operators and $\mu$ is the conformal dimension of the internal operator. When the external operator is the identity operator $\mathbf{1}$, the conformal block reduces to the Virasoro character of operator $\mu$,
\begin{equation}
\CF^{\mathbf{1}}_{c,\mu}(e^{2\pi i \tau})=\mathrm{ch}_{\mu}(e^{2\pi i \tau}).
\end{equation}
For convenience, $c$, $\lambda$ or $\mu$ will also be replaced with other labels of the model or the operator in the following context.

$\CF^{\lambda}_{c,\mu}(e^{2\pi i \tau})$ is non-zero only when the Verlinde coefficient $N^{\mu}_{\lambda\mu}$ is not zero. Given the model and external operator $\lambda$, the collection of all non-vanishing one-point conformal blocks $\{\CF^{\lambda}_{c,\mu}(e^{2\pi i \tau})\}$ transform among each other under the modular group $SL(2,\mathbb{Z})$, therefore form a representation of the modular group with dimension $\sum_{\mu}N^{\mu}_{\lambda\mu}$.

The modularity of one-point conformal block for minimal model was studied in \cite{2016arXiv161202134K}. Recall that the central charge for the $(p,q)$ model is
\begin{equation}
c_{p,q}=1-\frac{6(p-q)^2}{pq}.
\end{equation}
The irreducible modules of $(p,q)$ model are labeled by
\begin{equation}
\{(r,s)|1\leq r\leq q, 1\leq s\leq p\},
\end{equation}
with the conformal weight of the primary,
\begin{equation}\label{confweight}
h_{r,s}=\frac{(rp-sq)^2-(p-q)^2}{4pq}.
\end{equation}
Note that $(r,s)$ and $(q-r,p-s)$ label the same module because $h_{r,s}=h_{q-r,p-s}$, and irreducible modules are uniquely determined by their conformal weight. The vacuum module is labeled by $(1,1)$ or $(q-1,p-1)$ since $h_{1,1}=h_{q-1,p-1} = 0$.

Assume $p\geq3$ is an odd integer and without loss of generality $s\leq\frac{p-1}{2}$. If the external operator is the primary of the module $(r,s)$, $\CF^{(r,s)}_{(p,q), (m,n)}(e^{2\pi i \tau}) $\footnote{For convenience, the minimal model and the primary operator are represented by labels instead of central charge or conformal weight.} is non-vanishing for $(m,n)$ pairs,
\begin{equation}
\left\{(m,n)|\frac{r+1}{2}\leq m\leq q-\frac{r+1}{2},\,
\frac{p+1}{2}\leq n\leq p-\frac{s+1}{2} \right\},
\end{equation}
with the total number $S=\frac{(p-s)(q-r)}{2}$. Clearly, when $(r,s)=(1,1)$, $(m,n)$ runs over all irreducible modules as expected.

It is proved in \citep{2016arXiv161202134K} that for $(p,q)$ minimal model, given $(r,s)$ the $S=\frac{(p-s)(q-r)}{2}$ non-vanishing torus one-point conformal blocks form a holomorphic vector-valued modular form under modular $SL(2,\mathbb{Z})$. And the matrix representation for the modular $T$-transformation is an $s\times s$  diagonal matrix,
\begin{equation}
\label{eq:modularT1pt}
\CT^{(p,q)}_{(r,s)}
=\mathrm{diag}\{e^{2\pi i r_1},\ldots,e^{2\pi i r_s}\},
\end{equation}
with $r_j=h_{m_j,n_j}-\frac{c_{p,q}}{24}-\frac{h_{r,s}}{12}$. $\CS^{(p,q)}_{(r,s)}$ is then computed by constraints,
\begin{equation}
\label{eq:STconstraint}
(\CS\CT)^3=1,\,\,\,\,\,\CS^2=1.
\end{equation}
The modules will be always arranged in a way that $\CT^{(p,q)}_{(r,s)}$ reduces to $\CS$ for characters in Eq. \ref{eq:modularcha} when $(r,s)=(1,1)$. Given an arbitrary diagonal matrix $\CT$, the explicit form of its corresponding $\CS$ matrix was studied in \cite{Itoyama:2012re}. Their results at lower ranks are quoted in appendix \ref{sec:Smatrixsol}.

\section{Coulomb branch index of AD theories and torus one-point conformal of minimal models}
\label{sec:inter}

The goal of this section is to generalize Eq. \ref{eq:CBIandSTrelationMod}, which demonstrates the relationship between the Coulomb branch index of the $(A_1, A_{2N})$ AD theory and the modular property of $(2N+3,2)$, to an arbitrary value of $\fu$. To do this, a one parameter generalization of modular transformation matrices in Eq. \ref{eq:modularcha} is required, and this generalization can be obtained natually by looking at modular properties of torus one-point conformal blocks of $(2N+3,2)$ minimal models.

\subsection{Generalized modular transformation matrices from torus one-point conformal blocks}

As mentioned in section \ref{sec:torus1pt}, the torus one-point conformal blocks of $(2N+3,2)$ minimal models with the external state $(1, s)$ form a vector valued modular form of dimension $(2N+3-s)/2$, with the modular $T$-transformation given explicitly in Eq. \ref{eq:modularT1pt}. Given a series of vector valued modular forms with the same dimension, a one parameter family of $SL(2,\mathbb{Z})$ can be constructed, therefore can be viewed as a one parameter generalization of modular transformation matrices, Eq. \ref{eq:modularcha}. This will be demonstrated explicitly for lower dimensions first and then generalize to arbitrary dimensions.

\subsubsection*{Two dimensional representation}
\label{sec:twodim}

For $(2N+3,2)$ minimal models with $N$ being a positive integer, the torus one-point partition function \newline $\CF^{(r,s)}_{(2N+3,2),(m,n)}(e^{2\pi i \tau})$ forms a two dimensional representation under $SL(2,\mathbb{Z})$ if and only if the external module $(r,s)$ is labeled by $(1,2N-1)$, and non-vanishing internal modules are $(1,N+3)$ and $(1,N+2)$. These internal modules can also be labeled as $(1,N)$ and $(1,N+1)$ because of the doubling.

The matrix representation of the $T$-transformation is,
\begin{equation}
\label{eq:T2dim}
\CT^{(2N+3,2)}_{(1,2N-1)}
=e^{\frac{\pi i}{6}}
\left(
\begin{array}{cc}
e^{\frac{\pi i}{2N+3}} & 0 \\
0 & e^{-\frac{\pi i}{2N+3}}
\end{array}
\right).
\end{equation}
The matrix representation of the $S$-transformation can therefore be obtained by solving the constraint Eq. \ref{eq:STconstraint}. They reduce to modular transformation matrices of characters of $(5,2)$ minimal model when $m=1$.

It is easy to check that Eq. \ref{eq:T2dim} is just specialization of matrices $\CS_{(5,2)}(\fu)$ and $\CT_{(5,2)}(\fu)$,
\begin{equation}
\label{eq:quant2dim}
\begin{split}
\CT_{(5,2)}(\fu)
&=e^{\frac{\pi i}{6}}
\left(
\begin{array}{cc}
\fu^{-1/2} & 0 \\
0 & \fu^{1/2}
\end{array}
\right),\\
\CS_{(5,2)}(\fu)
&=\frac{1}{1-\fu}
\left(
\begin{array}{cc}
-i\fu^{1/2} & \sqrt{1-\fu+\fu^2} \\
\sqrt{1-\fu+\fu^2} & i\fu^{1/2}
\end{array}\right),
\end{split}
\end{equation}
when $\fu$ is set to be $e^{-\frac{2\pi i}{2N+3}}$:
\begin{equation}\label{eq:newid}
\begin{split}
\CT_{(5,2)}(\fu=e^{-\frac{2\pi i}{2N+3}})
=&\CT^{(2N+3,2)}_{(1,2N-1)}, \\
\CS_{(5,2)}(\fu=e^{-\frac{2\pi i}{2N+3}})
=&\CS^{(2N+3,2)}_{(1,2N-1)}.
\end{split}
\end{equation}
$\CS_{(5,2)}(\fu)$ and $\CT_{(5,2)}(\fu)$ satisfy the constraint Eq. \ref{eq:CBIA1A2},
\begin{equation}
(\CS_{(5,2)}(\fu)\CT_{(5,2)}(\fu))^3=1,\,\,\,\,\,(\CS_{(5,2)}(\fu))^2=1,
\end{equation}
for arbitrary $\fu$. These matrices form a one parameter family of two dimensional representation of $SL(2,\mathbb{Z})$, and can be viewed as a deformation of modular transformation matrices of characters of $(5,2)$ minimal model.

\subsubsection*{Three dimensional representation}

For $(2N+5,2)$ minimal models with $N$ being a positive integer, the torus one-point partition function \newline $\CF^{(r,s)}_{(2N+5,2),(m,n)}(e^{2\pi i \tau})$ forms a three dimensional representation under $SL(2,\mathbb{Z})$ if and only if the external module is $(1,2N-3)$. These series of $\CS^{(2N+5,2)}_{(1,2N-1)}$ and $\CT^{(2N+5,2)}_{(1,2N-1)}$ can be considered as the deformation of modular $S$ and $T$ matrices for the $(7,2)$ model, and are specialization of
\begin{equation}
\label{eq:ST72u}
\begin{split}
\CT_{(7,2)}(\fu)&=e^{\frac{\pi i}{3}}
\left(
\begin{array}{ccc}
\fu^{-\frac{5}{3}} & 0 & 0\\
0 & \fu^{\frac{1}{3}} & 0\\
0 & 0 & \fu^{\frac{4}{3}}
\end{array}
\right),\\
\CS_{(7,2)}(\fu)&=
\left(
\begin{array}{ccc}
 -\frac{\fu^2}{(\fu-1)^2 \left(\fu^2+\fu+1\right)} & -\frac{\sqrt{\fu^2+\fu}\sqrt{\fu^5+1}}{\sqrt{\fu-1} \left(\fu^2-1\right)
   \sqrt{\fu^3-1}} & -\frac{\sqrt{\fu^4+1} \sqrt{\fu^5+1}}{\sqrt{\fu-1} \sqrt{\fu^2-1} \left(\fu^3-1\right)} \\
 \frac{\sqrt{\fu^2+u}  \sqrt{\fu^5+1}}{\sqrt{\fu-1} \left(\fu^2-1\right) \sqrt{\fu^3-1}} & 1+\frac{1}{\fu-1}+\frac{1}{(\fu-1)^2}
   & \frac{\sqrt{\fu^4+1} \sqrt{\fu^2+\fu}}{(\fu-1) \sqrt{\fu^2-1} \sqrt{\fu^3-1}} \\
 -\frac{\sqrt{\fu^4+1} \sqrt{\fu^5+1}}{\sqrt{\fu-1} \sqrt{\fu^2-1} \left(\fu^3-1\right)} & -\frac{\sqrt{\fu^4+1}
   \sqrt{\fu^2+\fu}}{(\fu-1) \sqrt{\fu^2-1} \sqrt{\fu^3-1}} & -\frac{\fu \left(\fu^2+1\right)}{(\fu-1)^2 \left(\fu^2+\fu+1\right)}
   \\
\end{array}
\right).
\end{split}
\end{equation}
It is easy to check directly that,
\begin{equation}
(\CS_{(7,2)}(\fu)\CT_{(7,2)}(\fu))^3=1,\,\,\,\,\,(\CS_{(7,2)}(\fu))^2=1,
\end{equation}
for arbitrary $\fu$, and
\begin{equation}
\CT_{(7,2)}(e^{-\frac{2\pi i}{2N+5}})
=\CT^{(2N+5,2)}_{(1,2N-1)}.
\end{equation}

\subsubsection*{Arbitrary dimension}

In general the one parameter generalization of modular transformation matrices for $(2N+3,2)$ model with positive integer $N$ can be constructed by looking at the series of modular transformation matrices $\CS^{(2N+3+M,2)}_{(1,M+1)}$ and $\CT^{(2N+3+M,2)}_{(1,M+1)}$. The generalized $\CT_{(2N+3,2)}(\fu)$ matirx is diagonal with non-zero elements,
\begin{equation}
\label{eq:STN2}
\left(\CT_{(2N+3,2)}(\fu)\right)_{ii}
=e^{\frac{\pi i N}{6}}\fu^{-\frac{1}{6}N(2N+1)+\frac{2N+1}{2}i-\frac{i^2}{2}}.
\end{equation}
Note that the matrix index $i$ is chosen to run from $0$ to $N$ for later convenience, therefore $\left(\CT_{(2N+3,2)}(\fu)\right)_{00}$ is the vacuum-vacuum component of $\CT_{(2N+3,2)}(\fu)$.

In principle the generalized $\CS_{(2N+3,2)}$ can be solved using the constraint equations. For lower dimension  explicit expressions of $\CS_{(2N+3,2)}$ are summarized in the appendix \ref{sec:Smatrixsol}. Another way to obtain $\CS_{(2N+3,2)}$ is presented in section \ref{sec:rCS}.

\subsection{Coulomb branch indices as generalized modular transformation matrices}

With the help of $\CS_{(2N+3,2)}(\fu)$ and $\CT_{(2N+3,2)}(\fu)$, it is now natural to generalize the relation in Eq. \ref{eq:CBIandSTrelationMod} between Coulomb branch indices and modular properties of characters to arbitrary parameter $\fu$. Again, the relation will be checked explicitly for $(A_1, A_2)$ and $(A_1, A_{4})$ case and then generalized to $(A_1, A_{2N})$ cases.

\subsubsection*{$(A_1,A_2)$ case}
Starting with $(A_1, A_2)$ AD theory, its Couloub index is
\begin{equation}
\label{eq:CBIA1A2}
\CI_{(A_1,A_{2})}(\fu)=\frac{1}{(1-\fu^3)(1-\fu^2)}+
\frac{\fu^k}{(1-\fu^6)(1-\fu^{-1})},
\end{equation}
%and under the limit $\fu\rightarrow e^{\frac{2\pi i }{5}}$ the index $\CI_{(A_1,A_{2})}(\fu)$ goes to $e^{\pi i k/30}\left[
%\CS^{(5,2)}_{(1,1)}(\CT^{(5,2)}_{(1,1)})^k\CS^{(5,2)}_{(1,1)}\right]_{0,0}$, where $\CS^{(5,2)}_{(1,1)}$ and $\CT^{(5,2)}_{(1,1)}$ are the modular transformation matrices for torus one-point conformal blocks and in this case reduce to modular transformation matrices for characters.

%For general $(2N+3,2)$ model with $m$ being a positive integer, $\CS^{(2N+3,2)}_{(r,s)}$ and $\CT^{(2N+3,2)}_{(r,s)}$ are $2\times2$ matrices only when $(r,s)=(1,2m-1)$. Therefore the generalized relation between Coulomb index of $(A_1, A_2)$ theory and the modularity of $(2N+3,2)$ models is,
%\begin{equation}
%\lim_{\fu\rightarrow e^{\frac{2\pi i}{2N+3}}} \CI_{(A_1,A_2)}(\fu)
%=\frac{1-e^{\frac{2\pi i}{2N+3}}}{1-e^{\frac{12\pi i}{2N+3}}}e^{\left(\frac{1}{2N+3}-\frac{1}{6}\right)i\pi k}\left[
%\CS^{(2N+3,2)}_{(1,2m-1)}\left(\CT^{(2N+3,2)}_{(1,2m-1)}\right)^k\CS^{(2N+3,2)}_{(1,2m-1)}\right]_{0,0},
%\end{equation}
%with $m$ being a positive integer. This relation can be checked explicitly using the explicit formular of the index \ref{eq:CBIA1A2} and the modular matrices \ref{eq:modularT1pt}.

It is natual to ask if the Coulomb branch index $\CI_{(A_1,A_{2})}(\fu)$ is further related to $S_{(5,2)}(\fu)$ and $T_{(5,2)}(\fu)$. Explicit computation using Eq.'s \ref{eq:quant2dim} and \ref{eq:CBIA1A2}  tells us,
\begin{equation}
\label{eq:CBIA1A2}
\begin{split}
\CI_{(A_1,A_2)}(\fu)
&=e^{\frac{\pi i k}{6}}\fu^{\frac{k}{2}}\frac{1-\fu}{1-\fu^6}
\left[\CS_{(5,2)}^{-1}(\fu)\CT_{(5,2)}^{-k}(\fu)\CS_{(5,2)}(\fu)\right]_{0,0}\\
&=\fu^{k}\frac{1-\fu}{1-\fu^6}\left[\CS_{(5,2)}^{-1}(\fu)\CT_{(5,2)}^{-k}(\fu)\CS_{(5,2)}(\fu)\CT_{(5,2)}^k(\fu)\right]_{0,0}
,
\end{split}
\end{equation}
and the matrix index $0$ represents the vacuum module $(1,1)$. %{\color{red}The prefactor is determined by the ratio between the index on $S^3\times S^1$ and $S^{-1}T^{-1}S$.}
This is the most general relation between Coulomb branch index of $(A_1, A_2)$ theory on $L(k,1)\times S^1$ and the generalized modular transformation matrices the $(5,2)$ model, which encodes the modular properties of torus one-point conformal blocks of $(2N+3,2)$ models with $N\geq 1$.

\textbf{Remark:} Note again that the identification of $\fu$ is slightly modified from the relation in \cite{Fredrickson:2017yka} in order to match the refined Chern-Simons theory in the next section. In Eq. \ref{eq:CBIandSTrelation} as in the original work of \cite{Fredrickson:2017yka}, the limit is taken to be $\fu=e^{\frac{2\pi i}{2N+3}}$, whereas  $\fu=e^{-\frac{2\pi i}{2N+3}}$ in Eq.'s \ref{eq:CBIandSTrelationMod} and \ref{eq:newid}. The extra framing factor $T^k_{(5,2)}(\fu)$ may be introduced to remove the phase factor in the first line of Eq. \ref{eq:CBIA1A2}.

\subsubsection*{$(A_1,A_4)$ case}
The next one is $(A_1, A_4)$ AD theory with Coulomb index,
\begin{equation}
\begin{split}
\CI_{(A_1,A_4)}(\fu)&=\frac{1}{(1-\fu^{3})(1-\fu^4)(1-\fu^5)(1-\fu^2)}
+\frac{\fu^{k}}{(1-\fu^8)(1-\fu^{-1})(1-\fu^5)(1-\fu^2)}\\
&+\frac{\fu^{3k}}{(1-\fu^8)(1-\fu^{-1})(1-\fu^{10})(1-\fu^{-3})}.
\end{split}
\end{equation}
The complete relation with generalized modular transformation matrices of the $(7,2)$ model is
\begin{equation}
\begin{split}
\CI_{(A_1,A_4)}(\fu)
&=e^{\frac{\pi i k}{3}}\fu^{\frac{4k}{3}}\frac{(1-\fu)(1-\fu^3)}{(1-\fu^8)(1-\fu^{10})}
\left[\CS^{-1}_{(7,2)}(\fu)\CT_{(7,2)}^{-k}(\fu)\CS_{(7,2)}(\fu)\right]_{0,0}\\
&=\fu^{3k}\frac{(1-\fu)(1-\fu^3)}{(1-\fu^8)(1-\fu^{10})}
\left[\CS^{-1}_{(7,2)}(\fu)\CT_{(7,2)}^{-k}(\fu)\CS_{(7,2)}(\fu)\CT^k_{(7,2)}(\fu)\right]_{0,0}.
\end{split}
\end{equation}

\subsubsection*{$(A_1, A_{2N})$ case}
For general $(A_1, A_{2N})$ Argyres-Douglas theories on $L(k,1)\times S^1$, we conjecture that the Coulomb branch index should be able to expressed by the generalized modular transformation matrices of the $(2N+3,2)$ model,
\begin{equation}
\label{eq:CBIA1A2n}
\begin{split}
\CI_{(A_1,A_{2N})}=&
e^{\frac{\pi i Nk}{6}}\fu^{\frac{N(N+2)k}{6}}
\prod_{i=1}^N\frac{1-\fu^{2i-1}}{1-\fu^{2i+2N+2}}
\left[\CS^{-1}_{(2N+3,2)}(\fu)\CT^{-k}_{(2N+3,2)}(\fu)\CS_{(2N+3,2)}(\fu)\right]_{0,0}\\
=&
\fu^{\frac{N(N+1)k}{2}}
\prod_{i=1}^N\frac{1-\fu^{2i-1}}{1-\fu^{2i+2N+2}}
\left[\CS^{-1}_{(2N+3,2)}(\fu)\CT^{-k}_{(2N+3,2)}(\fu)\CS_{(2N+3,2)}(\fu)\CT^{k}_{(2N+3,2)}(\fu)\right]_{0,0}.
\end{split}
\end{equation}
This relation has been checked explicitly up to $N=5$ case. It would be nice to have a proof of this conjecture, which may require better understanding of the generalized modular transformation matrices in Eq.'s \ref{eq:quant2dim}, \ref{eq:ST72u} and \ref{eq:STN2}.

In general, the Coulomb branch indices of $(A_1, A_{2N})$ theories on $L(k,1)\times S^1$ is proportional to the $(0,0)$ component of $\CS^{-1}_{(2N+3,2)}\CT^{-k}_{(2N+3,2)}\CS_{(2N+3,2)}$ up to a normalization factor and possible framings, where generalized modular transformation matrices $\CS_{(2N+3,2)}(\fu)$ and $\CT_{(2N+3,2)}(\fu)$ encode the modular properties of $(2M+3,2)$ models with $M\geq N$. Due to the fact that $\CS^2_{(2N+3,2)}=1$, the difference between $\CS^{-1}_{(2N+3,2)}\CT^{-k}_{(2N+3,2)}\CS_{(2N+3,2)}$ and \newline $\CS_{(2N+3,2)}\CT^{-k}_{(2N+3,2)}\CS_{(2N+3,2)}$ can not to be seen in this setup.

\section{Relation with refined Chern-Simons}
\label{sec:rCS}

As mentioned in the introduction, due to similar M-theory constructions, $(A_1, A_{2N})$ AD theories are expected to related to the refined CS theory. It is then important to understand this relation from the geometry first.

The $(A_1, A_{2N})$ AD theories are constructed by compactifying M5 branes on  a sphere $\Sigma_{N}$ with one irregular singularity. The Higgs field of the corresponding Hitchin system has the asymptotic behavior
\begin{equation}
\label{eq:A1A2NHiggs}
\Phi(z)dz\sim z^{\frac{2N+1}{2}}\sigma^3dz,
\end{equation}
where $z$ is the coordinate of the disk, and $\sigma^3$ is the third Pauli matrix. The singularity is placed at the infinity. There is one $\mathbb{C}^\ast$ action on the disk and another $\mathbb{C}^\ast$ action on the Higgs bundle. To compute the Coulomb branch indices, the AD theories are further placed on $L(k,1)\times S^1$, therefore the M5 branes are on $L(k,1)\times S^1\times \Sigma_N$.

Now recall the construction of $SU(N)$ refined CS theory \cite{Aganagic:2011sg}. Consider the M-theory on
\begin{equation}
(T^{\ast}M\times TN\times S^1)_{q,t},
\end{equation}
where $T^{\ast}M$ is the cotangent bundle of a three-manifold $M$ and $TN$ is the Taub-NUT space twisted along the $S^1$. The twisting is defined such that going around the $S^1$ circle, the complex coordinates ($z_1$, $z_2$) of the $TN$ rotate by
\begin{equation}
z_1\mapsto q z_1,\,\,\,\,\,z_2\mapsto t^{-1}z_2.
\end{equation}
One may add $N$ M5 branes wrapping
\begin{equation}
(M\times\mathbb{C}_{z_1}\times S^1)_q,
\end{equation}
where $M$ is the previous three-manifold, and $\mathbb{C}_{z_1}$ is the subspace of $TN$ parametrized by $z_1$. The refined CS partition function is defined as the corresponding M5 partition function,
\begin{equation}
Z^{rCS}(M,\rq,\rt)\equiv Z_M(T^{\ast}M,\rq,\rt).
\end{equation}

It is then natural to identify the $\mathbb{C}_{z_1}$ with $\Sigma_N$ of AD theories and the rotation on $\mathbb{C}_{z_2}$ with the $U(1)$ action on the Higgs bundle. However, the rotation around $z_1$ and $z_2$ can not be arbitrary otherwise the Higgs field in Eq. \ref{eq:A1A2NHiggs} will not be invariant. Going around the $S^1$ circle, the Higgs field $\Phi(z)$ becomes,
\begin{equation}
\tilde{\Phi}(\tilde{z})
=\rt\Phi(\rq z) = \rt \rq^{\frac{2N+1}{2}}\Phi(z),
\end{equation}
therefore the invariance of Higgs field requires that $\rt \rq^{\frac{2N+1}{2}}=1$, or
\begin{equation}
\label{eq:qtconstraints}
\rt^2\rq^{2N+1}=1.
\end{equation}
Moreover, the Coulomb branch indices of $(A_1, A_{2N})$ AD theories on $L(k,1)\times S^1$ is expected to be equal to the refined CS partition function on $L(k,1)$ with $\rt^2\rq^{2N+1}=1$ up to a normalization factor. This will be shown by explicit computation in the next sections.

Note that there is another construction of $(A_1, A_{2N})$ theories by considering type IIB string theory on isolated singularities in $\mathbb{C}^4$ defined by a polynomial \cite{Wang:2016yha},
\begin{equation}
x^2+y^2+z^{2N+1}+w^2=0.
\end{equation}
It would be interesting to understand the relation with refined CS theory via this construction, but it will not be the subject of this work.

\subsection{Refined CS representation}
\label{sec:rCSreps}

The refined $SU(2)_K$ CS topological quantum field theory (TQFT)\footnote{The CS level is denoted by $K$ to avoid confusion with $k$ in $L(k,1)$.} representations of mapping class groups of genus $1$ surface is summarized here. Comparing to the normal $SU(2)_K$ CS TQFT, the Hilbert space is unchanged but the matrix elements of generators $S$ and $T$ depends on two parameters $\rq$ and $\rt$ \cite{Arthamonov:2015rha},
\begin{eqnarray}
\label{ea:repMCGg1}
\langle i|T|j\rangle&\equiv& T_i(\rq,\rt)\delta_{ij}
= \rq^{-j^2/4}\rt^{-j/2}\delta_{ij},\\
\langle i|S|j\rangle&\equiv& S_{ij}(\rq,\rt)
=S_{00} q^{ij/2}g^{-1}_i P_i(\rt^{\half},\rt^{-\half};\rq,\rt)
P_j(\rt^{\half}\rq^i,\rt^{-\half};\rq,\rt),
\end{eqnarray}
where $S_{00}$ is a normalization constant, and $i$ and $j$ run over non negative integers and are the Dynkin label of $SU(2)$ irreducible representations. $P_j(x_1,x_2;\rq,\rt)$ is the $SU(2)$ Macdonald polynomial of the spin-$j/2$ representation, and $g_j$ is the quadratic norm of the Macdonald polynomials $P_j(x_1,x_2;\rq,\rt)$ under a natural orthogonality condition. The explicit forms and properties of $P_j(x_1,x_2;\rq,\rt)$ and $g_j$ are summarized in appendix \ref{app:sec:MDpoly}.
$q$ and $t$ are related to the CS level $K$ by the set of relations $\rq=e^{\frac{2\pi i}{K+2\beta}}$, $\rt=\rq^{\beta}=e^{\frac{2\pi i\beta}{K+2\beta}}$, but we will not use this relation in our paper.

The refined operators satisfy the same $SL(2,\Z)$ relations $S^2=1$ and $(ST)^3 \propto \mbox{id}$, and they reduce to the usual CS operators when $t=q$ ($\beta=1$). $T_i$ and $S_{ij}$ have infinitely many components in general. However, if $q$ and $t$ satisfies the following relations,
\begin{equation}
q^nt^2=1,\,\forall\ n\in\mathbb{Z},\,n\geq0,
\end{equation}
the Macdonald polynomial at $(x_1=\rt^\half,x_2=\rt^{-\half})$,\footnote{Also called the ($\rq,\rt$)-deformed dimension of spin-$j/2$ representation.} $P_j(\rt^\half,\rt^{-\half};\rq,\rt)$ vanishes for $j>n$,
\begin{equation}
P_j(\rt^\half,\rt^{-\half};\rq,\rt) = 0, \,\,\,\,\,\forall\ j>k.
\end{equation}
Hence, $S_{ij}$ is truncated to a $(n+1)$ by $(n+1)$ matrices when $\rq^n\rt^2=1$, and only the first $n+1$ entries of $T_i$ are relevant here.

For a three-manifold $M$ constructed by gluing the boundaries of two solid tori up to an $SL(2,\mathbb{Z})$ transformation $V(q,t)$, the refined CS partition function on $M$ is,
\begin{equation}
Z^{rCS}(M;q,t)=\langle0|V(q,t)|0\rangle,
\end{equation}
where $V(q,t)$ is the refined CS representation of $M$.

\subsection{Generalized modular matrices of minimal models and refined CS representations}
\label{sec:matching2dim}

Equation \ref{ea:repMCGg1}, which is the refined CS representation of $T$-transformation, matches with quantized $\CT_{(2N+3,2)}(\fu)$ (Eq. \ref{eq:modularT1pt}) up to an overall constant  under the change of variables,
\begin{equation}
\label{eq:CStoCBILimit}
\begin{split}
\rq&\rightarrow \fu^{2}, \\
\rt&\rightarrow \fu^{-2N-1}.
\end{split}
\end{equation}
To be precise,
\begin{equation}
\left(\CT_{(2N+3,2)}(\fu)\right)_{ii}
=e^{\frac{\pi i N}{6}}\fu^{-\frac{1}{6}N(2N+1)}T_i(\fu^{2},u^{-2N-1}),\,\,\,\,\,0\leq i\leq N.
\end{equation}
under limit in Eq. \ref{eq:CStoCBILimit}, $\rq^{2N+1}\rt^2=\fu^{-2(2N+1)}\fu^{-(2N+1)2}=1$. Hence, $S_{ij}(\rq,\rt)$ and $T_i(\rq,\rt)$ are truncated to $0\leq i\leq 2N+1$, and act on a $2N+2$ dimensional linear space. It will be shown that the actual Hilbert space is $N+1$ dimensional!

There is a symmetry in $T_i(\fu^{2},\fu^{-2N-1})$,
\begin{equation}
T_i(\fu^{2},\fu^{-2N-1})=T_{2N+1-i}(\fu^{2},\fu^{-2N-1}),
\,\,\,\,\,0\leq i\leq 2N+1,
\end{equation}
therefore it is natural to identify the $2N+2$ dimensional Hilbert space $\{|i\rangle|0\leq i\leq 2N+1\}$, on which operator $T$ acts, with the space of irreducible modules $\{(1,n)|1\leq n\leq 2N+2\}$, and the symmetry in $T_i(\fu^{2},\fu^{-2N-1})$ is interpreted as the identification of $(1,i+1)$ module and $(1,2m-i+1)$ module. This identification is further supported by the observation that only half of eigenvalues of $S_{ij}(\fu^{2},\fu^{-2N-1})$ are zero, hence the non-trivial eigenspace of $S_{ij}(\fu^{2},\fu^{-2N-1})$ is only $N+1$ dimensional, coinciding with the number of irreducible modules of $(2N+3,2)$ model.

\subsubsection*{Two dimensional case}

When $N=1$ everything can be worked out explicitly. After substituting $\rq=\fu^{2}$ and $\rt=\fu^{-3}$ the matrix representation of the $T$ operator is,
\begin{equation}
T_i(\fu^{2},\fu^{-3})\delta_{ij}
=\left(
\begin{array}{cccc}
1 &&&\\
&\fu&&\\
&&\fu&\\
&&&1
\end{array}
\right),
\end{equation}
and the representation for $S$ operator is
\begin{equation}
\begin{split}
\label{eq:Srep2dim}
& S_{ij}(\fu^2,\fu^{-3})
= \\ & S_{00}\left(
\begin{array}{cccc}
 1 & \frac{u^3+1}{u^{3/2}} & 2+\frac{\left(u^2+u+1\right) \left(u^4+1\right)}{u^3} & -\frac{(u+1)^3 ((u-1) u+1) \left(u^2+1\right)}{u^{7/2}} \\
 -\frac{\sqrt{u}}{u+1} & -1 & -\frac{(u+1) \left(u^2+1\right)}{u^{3/2}} &2+ \frac{\left(u^2+u+1\right) \left(u^4+1\right)}{u^3} \\
 -\frac{u^2}{(u+1)^2 \left(u^2+1\right)} & -\frac{u^{3/2}}{u^3+u^2+u+1} & -1 & \frac{u^3+1}{u^{3/2}} \\
 -\frac{u^{7/2}}{(u+1)^3 ((u-1) u+1) \left(u^2+1\right)} & -\frac{u^2}{(u+1)^2 \left(u^2+1\right)} & -\frac{\sqrt{u}}{u+1} & 1 \\
\end{array}
\right).
\end{split}
\end{equation}
Entries with value $0$ are omitted in the above expressions.

To match with $\CT_{(5,2)}(\fu)$ and $\CS_{(5,2)}(\fu)$, one perform a similarity transformation such that $S$ becomes block diagonal with only upper left block none zero and $T$ remains the same,
\begin{equation}
\label{eq:SS2dim}
S'_{ij}(\fu^{2},\fu^{-3})= \Omega_1^{-1}\,S_{ij}(\fu^{2},\fu^{-3})\,\Omega_1
=\frac{2i}{\sqrt{\fu}}S_{00}\left(
\begin{array}{cccc}
-i\sqrt{\fu} & \sqrt{1-\fu+\fu^2} &&\\
\sqrt{1-\fu+\fu^2} & i\sqrt{\fu}&&\\
&& 0 & 0 \\
&& 0  & 0
\end{array}\right),
\end{equation}
and
\begin{equation}
\label{eq:TT2dim}
T'_{ij}(\fu^{2},\fu^{-3})= \Omega_1^{-1}\,T_i(\fu^{2},\fu^{-3})\delta_{ij}\,\Omega_1
=\left(
\begin{array}{cccc}
1 &&&\\
&\fu&&\\
&&\fu&\\
&&&1
\end{array}
\right).
\end{equation}
The $(i,j)$ entry of the transformation matrix $(\Omega_1)_{ij}$ is non-zero only when $i=j$ or $i=2N+1-j$ to keep $T'$ the same as $T$. The explicit derivation of $\Omega_1$ is left in the appendix \ref{app:sec:similarity}.

Denoting the upper-left diagonal blocks of $S'_{ij}(\fu^{2},\fu^{-3})$ and $T'_{ij}(\fu^{2},\fu^{-3})$ by $S^{r}_1(\fu)$ and $T^{r}_1(\fu)$ respectively, one obtains
\begin{equation}
\label{eq:matchingST2dim}
\begin{split}
S^{r}_1(\fu)&=\frac{2i(1-\fu)}{\sqrt{\fu}}S_{00}\CS_{(5,2)}(\fu),\\
T^{r}_1(\fu)&=e^{-\frac{\pi i}{6}}\fu^{\half}\CT_{(5,2)}(\fu).
\end{split}
\end{equation}
Therefore with a suitable rotation of basis and the constraint $q^2t^3=1$, the refined $SU(2)$ CS representation of mapping class group matches with the quantized $\CS$ and $\CT$ of the $(5,2)$ minimal model.

%It follows from the definition of $S^r_1$ and $T^r_1$ that
%\begin{equation}
%\begin{split}
%(A_1)_{00}(S^r_1(T^r_1)^kS^r_1)_{00}(A^{-1}_1)_{00}
%=&\sum_{i,j}(A_1)_{0i}(S'(T')^kS')_{ij}(A^{-1}_1)_{j0}\\
%=&\sum_{i}(S_{0i} T^k_iS_{i0}).
%\end{split}
%\end{equation}
%The first equality is because that $(A_2)_{0i}$ is non zero only when $i=0$ or $i=2N+1$ and $(S'(T')^kS')_{ij}$ is zero when $i>N$ or $j>N$. Using the fact that $(A_2)_{00}(A^{-1}_2)_{00}=\half$ and equation \ref{eq:matchingST2dim},
%\begin{equation}
%(\CS_{(5,2)}\CT^k_{(5,2)}\CS_{(5,2)})_{00}
%=-\frac{\fu}{2(1-\fu)^2}e^{\pi ik/6}\fu^{k/2}
% \frac{\sum_{i}S_{0i}T_{i}^kS_{i0}}{S_{00}^2}.
%\end{equation}
%Further applying equation \ref{eq:CBIA1A2},
%\begin{equation}
%\begin{split}
%\CI_{(A_1,A_2)}(\fu)&=-\frac{\fu^{k+1}}{2(1-\fu)(1-\fu^6)}
%\frac{\sum_{i}S_{0i}T^k_iS_{i0}}{S^2_{00}}\\
%&=-\frac{\fu^{k+1}}{2(1-\fu)(1-\fu^6)}\frac{1}{S^2_{00}}
%Z^{rCS}(S^3/\Z_k;\fu^{-2},\fu^{3}).
%\end{split}
%\end{equation}
%Hence the Coulomb branch limit of the lens space $S^3/\Z_k$ index of $(A_1,A_2)$ AD theory is the refined CS partition function on the same lens space $S^3/\Z_k$ up to a proportional factor.

\subsubsection*{Arbitrary dimension}

The strategy to match $S_{ij}(\rq,\rt)$ and $T_i(\rq,\rt)$ operators in refined CS representation at $\rq=\fu^{2}$ and $\rt=\fu^{-2N-1}$ with $\CS_{(2N+3,2)}$ and $\CT_{(2N+3,2)}$. Again using the fact that the non-zero eigenspace of $S_{ij}(\fu^{2},\fu^{-2N-1})$ is $N+1$ dimensional instead of $2N+2$ dimensional, one can find a similarity transformation $\Omega_N$ which keeps $T_i(\fu^{2},\fu^{-2N-1})$ invariant but rotates $S_{ij}(\fu^{2},\fu^{-2N-1})$ such that only the $(N+1)$ by $(N+1)$  upper left block of $S_{ij}(\fu^{2},\fu^{-2N-1})$ is non-zero. Similar to $N=1$ case, define,
\begin{equation}
\begin{split}
S^r_{N}(\fu)&=[\Omega^{-1}_N \,S_{ij}(\fu^{2},\fu^{-2N-1})\, \Omega_N]_{(N+1)\times(N+1)},\\
T^r_{N}(\fu)&=[\Omega^{-1}_N \,T_i(\fu^{2},\fu^{-2N-1})\delta_{ij}\, \Omega_N]_{(N+1)\times(N+1)},
\end{split}
\end{equation}
where $[M]_{(N+1)\times(N+1)}$ means keeping only the $(N+1)$ by $(N+1)$  upper left block of the matrix $M$. By definition $S^r_N(\fu)$ and $T^r_N(\fu)$ satisfy the $SL(2,\Z)$ constraints up to normalization, and  are proportional to $\CS_{(2N+3,2)}(\fu)$ and $\CT_{(2N+3,2)}(\fu)$ up to an overall factor,
\begin{equation}
\label{eq:matchingNdim}
\begin{split}
\frac{S^r_N(\fu)}{S^{-1}_{00}}&=2e^{\frac{\pi iN}{2}}\frac{\prod_{i=1}^N(1-\fu^{2i-1})}{\fu^{N^2/2}}\CS_{(2N+3,2)}(\fu)\\
T^r_N(\fu)&=e^{-\frac{\pi iN}{6}}\fu^{\frac{1}{6}N(2N+1)}
\CT_{(2N+3,2)}(\fu).
\end{split}
\end{equation}
Therefore, the modular transformation matrices of intertwiners of $(2N+3,2)$ minimal models are mapped to the refined CS representation of mapping class group of torus with $\rq=\fu^{-2}$ and $\rt=\fu^{2N+1}$. Eq. \ref{eq:matchingNdim} provides another way to compute $\CS_{(2N+3,2)}(\fu)$ when $N$ is large.

\subsection{Coulomb branch indices and refined CS partition functions}
\label{sec:CBIandrCSpartitions}

It is explained in section \ref{sec:inter} that the Coulomb branch index $\CI_{(A_1, A_{2N})}(\fu)$ of $(A_1, A_{2N})$ AD theory on $L(k,1)\times S^1$ can be expressed as the combination of  $\CS_{(2N+3,2)}(\fu)$ and $\CT_{(2N+3,2)}(\fu)$ comes from the modular transformations of $(2N+3+m,2)$ minimal models. Using the result in the previous section, $\CS_{(2N+3,2)}(\fu)$ and $\CT_{(2N+3,2)}(\fu)$ are proportional to the refined CS representation of $S$ and $T$ operators of the mapping class group of the torus. Therefore $\CI_{(A_1, A_{2N})}$ can be identified with the refined CS partition function.

Using Eq.'s \ref{eq:CBIA1A2n} and \ref{eq:matchingNdim}, one expresses the Coulomb branch index of $(A_1, A_{2N})$ AD theory by matrices $S^r_N$ and $T^r_N$,
\begin{equation}
\begin{split}
\CI_{(A_1,A_{2N})}(\fu)=&
\fu^{\half N(N+1)k}\prod_{i=1}^N\frac{1-\fu^{2i-1}}
{(1-\fu^{2i+2N+2})}
\sum_{i}(S^r_N)^{-1}_{0i}(T^r_N)_i^{-k}(S_N^r)_{i0}\\
=&\fu^{\half N(N+1)k}\prod_{i=1}^N\frac{1-\fu^{2i-1}}
{(1-\fu^{2i+2N+2})}
\sum_{i}(S^r_N)^{-1}_{0i}(T^r_N)_i^{-k}(S_N^r)_{i0}(T^r_N)_0^{k}.
\end{split}
\end{equation}
The second line follows naturally from the fact that $(T^r_N)_0=1$.
%Using the fact that $(A_N)_{ij}$ are non-zero only when $i=j$ and $i=2N+1-j$ and $(A_N)_{00}(A^{-1}_N)_{00}=\half$,
%\begin{equation}
%\begin{split}
%\half\sum_{i}(S^r_N)_{0i}(T^r_N)_i^k(S_N^r)_{i0}
%&=
%\sum_{i}(A_{N})_{00}(S^r_N)_{0i}(T^r_N)_i^k(S_N^r)_{i0}(A^{-1}_N)_{00}\\
%&=\sum_{i}S_{0i}(\fu^{2},\fu^{-2N-1})
%T^k_{i}(\fu^{2},\fu^{-2N-1})S_{i0}(\fu^{2},\fu^{-2N-1}).
%\end{split}
%\end{equation}
In terms of refined CS theory,
\begin{equation}
\begin{split}
\CI_{(A_1,A_{2N})}(\fu)=&
(-1)^N\frac{\fu^{\half N(N+1)k+N^2}}
{2\prod_{i=1}^N(1-\fu^{2i-1})(1-\fu^{2i+2N+2})}\frac{1}{S^2_{00}}\\
&\times\sum_{i}S_{0i}(\fu^{2},\fu^{-2N-1})
T^{-k}_{i}(\fu^{2},\fu^{-2N-1})S_{i0}(\fu^{2},\fu^{-2N-1})\\
=&(-1)^N\frac{\fu^{\half N(N+1)k+N^2}}
{2\prod_{i=1}^N(1-\fu^{2i-1})(1-\fu^{2i+2N+2})}\frac{1}{S^2_{00}}\\
&\times\sum_{i}S_{0i}(\fu^{2},\fu^{-2N-1})
T^{-k}_{i}(\fu^{2},\fu^{-2N-1})S_{i0}(\fu^{2},\fu^{-2N-1})T^{k}_{0}(\fu^{2},\fu^{-2N-1}).
\end{split}
\end{equation}
$S(\fu^2,\fu^{-2N-1})$ is used here instead of $S^{-1}$ is because that under this specialization $S(\fu^2,\fu^{-2N-1})$ is singular and the inverse only exists in the $N+1$ dimensional subspace discussed before\footnote{Technically this changes the orientation of the manifold by gluing to solid tori, however, indices in this paper are not sensitive to the orientation because $\CS^2=1$.}.
Notice that $S_{00}$ is a normalization factor depends on $\rq$ and $\rt$. In order to scale the eigenvalues of $S(\fu^2,\fu^{-2N-1})$ to either $1$ or $0$, $S_{00}$ is chosen as,
\begin{equation}
S^2_{00}(\rq,\rt)=\half\frac{(q^{\half};q)_{\infty}(t^{-1};q)_{\infty}}
{(tq^{-\half};q)_{\infty}(t^2;q)_{\infty}},
\end{equation}
with the $q$-Pochhammer symbol $(a;q)_{\infty}\equiv\prod_{i=0}^{\infty}(1-aq^i)$. With this normalization factor it can be shown that
\begin{equation}
2(-1)^N \fu^{-N^2}\prod_{i=1}^N(1-\fu^{2i-1})(1-\fu^{2i+2N+2})S^2_{00}(\fu^{-2},\fu^{2N+1}) = 1.
\end{equation}

Recall that one construction of lens space $L(k,1)$ is by gluing two solid tori with an $SL(2,\mathbb{Z})$ transformation $S^{-1}T^{-k}S$ up to framing factors. Therefore the Coulomb branch of AD theories on $L(k,1)\times S^1$ should be identified with the refined CS partition function on $L(k,1)$. The relation between $\CI_{(A_1,A_{2N})}(\fu)$ and $Z^{rCS}(L(k,1);u^{-2},u^{2N+1})$ simplifies after  the above normalization
\begin{equation}
\CI_{(A_1,A_{2N})}(\fu)
=\fu^{\half N(N+1)k}
Z^{rCS}(L(k,1);\fu^{-2},\fu^{2N+1}).
\end{equation}
Therefore the Coulomb branch index of $(A_1, A_{2N})$ AD theory on $L(k,1)\times S^1$ is indeed the refined CS partition function on $L(k,1)$ up to an overall factor.

\section{Further Generalizations}
\label{sec:generalizations}

\subsection{Partition functions of AD theories on $L(p,q)\times S^1$}

It is easy to compute the refined CS partition function on general lens space $L(p,q)$. Written $p/q$ as a continued fraction $[a_0;a_1,a_2,\cdots,a_n]$\footnote{The definition of continued fraction in this paper is slightly different from the usual one.},
\begin{equation}
\frac{p}{q}=a_0-\frac{1}{a_1-\frac{1}{a_2-\frac{1}{\ddots-\frac{1}{a_n}}}},
\end{equation}
the gluing elements for $L(p,q)$ is then,
\begin{equation}
S^{-1}T^{-a_0}S^{-1}T^{-a_1}\cdots S^{-1}T^{-a_n}ST^{\sum_n a_n},
\end{equation}
where the last term $T^{\sum_n a_n}$ corresponds to a choice of framing.

The supersymmetric partition function of $(A_1, A_{2N})$ AD theories on $L(p,q)\times S^1$ is conjectured to be (again, using the fact that $\CS_{(2N+3,2)}^2=1$),
\begin{equation}
\begin{split}
\CI^{L(p,q)\times S^1}_{(A_1, A_{2N})}
&=
c_N(\fu)\left(\CS_{(2N+3,2)}\CT_{(2N+3,2)}^{-a_0}\cdots\CS_{(2N+3,2)}\CT_{(2N+3,2)}^{-a_n}\CS_{(2N+3,2)}\CT_{(2N+3,2)}^{\sum_na_n}\right)_{00}\\
&\propto Z^{rCS}(L(p,q);\fu^2,\fu^{-2N-1}),
\end{split}
\end{equation}
where $c_N(\fu)$ is a proportional factor which could depend on the zero-point energy of the partition function. It is interesting to compare this conjecture with direct localization computation and fix the ambiguity in zero-point energy and framing \cite{PYY2018}. This could be a potential way to compute the supersymmetric partition function of the $(A_1, A_{2N})$ AD theory on $M\times S^1$ with $M$ being a three manifold, and it would be nice to explore the possible relationship with other works on similar topics \cite{Gukov:2017kmk, Dedushenko:2017tdw, Gukov:2017zao, Moore:2017cmm}.

\subsection{Surface defects in AD theories and knot homology}
\label{eq:knothomology}

One natural object in refined CS theory is the Wilson line operator on a knot. In fact one remarkable application of the refined CS theory is to compute the knot homology of torus knots. In the AD theory side this line operators are lifted to surface defects wrapping torus knots and $S^1$ coming from boundaries of M2 branes ending on the M5 brane.
Again using the identification between AD theories and refined CS, it is reasonable to assume that the supersymmetric partition function of $(A_1, A_{2N})$ AD theories with these surface defects inserted is proportional to the refined CS partition function with Wilson lines and computed in a similar fashion.
%
%For a general torus knot $K$, the relation is
%\begin{equation}
%I^{(2N+3,2)}_1(\fu,K)=\fu I^{\mathrm{Kh}}_1(\fu^2,\fu^{-(2N+1)},K).
%\end{equation}
%The standard knot homology $I^{\mathrm{Kh}}_1(\fq,\ft,K)$ is the refined CS partition function on $S^3\backslash K$ normalizaed by the partition function on $S^3$, with $K$ interpreted as a line operator insertion. In the AD theory side we interpret this as one direction of the M2 brane and the other direction of the M2 brane inside M5 should wrap on $S^1$.
%\begin{equation}
%\CI_{(A_1,A_{2N})}(\fu,S^3\times S^1,K\times S^1)
%=I^{(2N+3,2)}_1(\fu,K)
%\CI_{(A_1,A_{2N})}(\fu,k=1).
%\end{equation}
%Hence we obtain a expression for the partition function of AD theory with a surface defects which is supported on $K\times S^1$ inserted.
%

To compute the the effect of the surface defect, one first define the Verlinde coefficients using $\CS_{(2N+3,2)}(\fu)$,
\begin{equation}
\left(N^{(2N+3,2)}(\fu)\right)_{ijk}\equiv
\sum_{l=0}^{N}
\frac{\left(\CS_{(2N+3,2)}(\fu)\right)_{li}\left(\CS_{(2N+3,2)}(\fu)\right)_{lj}\left(\CS_{(2N+3,2)}(\fu)\right)_{lk}}{\left(\CS_{(2N+3,2)}(\fu)\right)_{l0}},
\end{equation}
and $N^{(2N+3,2)}_{i}$ is defined as the matrix with the following entries,
\begin{equation}
\left(N^{(2N+3,2)}_{i}\right)_{jk}=\left(N_{(2N+3,2)}(\fu)\right)_{ijk}.
\end{equation}
Therefore the Poincare invariants for a torus knot $K$ is
\begin{equation}
P^{(2N+3,2)}_i(\fu,K)=\frac{\left(K_{(2N+3,2)}(\fu)N^{(2N+3,2)}_i(\fu)K^{-1}_{(2N+3,2)}(\fu)\CS_{(2N+3,2)}(\fu)\right)_{00}}{\left(N^{(2N+3,2)}_i(\fu)\CS_{(2N+3,2)}(\fu)\right)_{00}},
\end{equation}
where $K_{(2N+3,2)}(\fu)$ is the quantized representation of the $SL(2,\Z)$ transformation which takes $(1,0)$ cycle on a torus to the knot $K$. $P_0$ is always $1$ by definition and $P_1$ gives the specialization of the usual Poincare polynomial. The supersymmetric partition function of $(A_1, A_{2N})$ theory is then conjectured to be,
\begin{equation}
\CI_{(A_1, A_{2N})}(\fu,K\times S^1)
=\CI_{(A_1,A_{2N})}(\fu,k=1)P^{(2N+3,2)}_1(\fu, K).
\end{equation}
It is interesting to explore further the meaning of the subscript $i$ in AD theories.

Like the refined CS, the partition function $\CI_{(A_1, A_{2N})}(\fu,K\times S^1)$ is closely related to knot homology. Examples are provided below to illustrate the connection between knot homology and $P^{(2N+3,2)}_i(\fu)$.

\subsubsection*{Example: The trefoil knot}

The trefoil knot is also the $(2,3)$ cycle on the torus and the $SL(2,\Z)$ transformation is,
\begin{equation}
K_{23}=ST^{-2}ST^{-2}.
\end{equation}
Using the data from $(2N+3,2)$ models, one gets,
\begin{equation}
P^{(2N+3,2)}_1(\fu,K_{23})
= -\fu + \fu^{-2N} +\fu^{-4N+1}=\frac{\CI_{(A_1, A_{2N})}(\fu,K\times S^1)}{\CI_{(A_1, A_{2N})}(\fu,k=1)}.
\end{equation}
The standard Poincare polynomial for the trefoil knot is
\begin{equation}
\mathrm{Kh}(\fq,\ft,K_{23})=-1+\ft^{-1}+\fq^{-1}\ft^{-2}.
\end{equation}
It is clear that
\begin{equation}
P^{(2N+3,2)}_1(\fu,K_{23})
=\fu \,\mathrm{Kh}(\fu^{-2},\fu^{2N+1},K_{23}).
\end{equation}

%\subsubsection*{The $(2,5)$ knot}
%
%The $SL(2,\Z)$ transformation which maps $(1,0)$ cycle to $(2,5)$ cycle is,
%\begin{equation}
%K_{25}=ST^{-3}ST^{-2},
%\end{equation}
%and
%\begin{equation}
%I^{(2N+3,2)}_1(\fu,K_{25})
%=-\fu-\fu^{-2N+2}+\fu^{-2N}+\fu^{-4N+1}+\fu^{-6N+2}.
%\end{equation}
%The standard knot homology for the $(2,5)$ knot is
%\begin{equation}
%I_1^{\mathrm{Kh}}(\fq,\ft,K_{25})=-1-\ft^{-1}\fq^{-1}+\ft^{-1}+\ft^{-2}\fq^{-1}+\ft^{-3}\fq^{-2}.
%\end{equation}
%Hence,
%\begin{equation}
%I^{(2N+3,2)}_1(\fu,K_{25})
%=\fu I^{\mathrm{Kh}}_1(\fu^2,\fu^{-(2N+1)},K_{25}).
%\end{equation}

\section{Conclusions and discussions}

The main conclusion of this paper is the relation among the Coulomb branch index of the $(A_1, A_{2N})$ AD theory on $L(k,1)\times S^1$, generalized modular transformation matrices of the $(2N+3,2)$ minimal model and the partition function of refined CS theory on $L(k,1)$,
\begin{equation}
\begin{split}
\CI_{(A_1,A_{2N})}
=&
\fu^{\half N(N+1)k}
\prod_{i=1}^N\frac{1-\fu^{2i-1}}{1-\fu^{2i+2N+2}}
\left[\CS^{-1}_{(2N+3,2)}(\fu)\CT^{-k}_{(2N+3,2)}(\fu)\CS_{(2N+3,2)}(\fu)\CT^{k}_{(2N+3,2)}(\fu)\right]_{0,0}\\
=&\fu^{\half N(N+1)k}
Z^{rCS}(L(k,1);\fu^{-2},\fu^{2N+1}),
\end{split}
\end{equation}
where $\CS_{(2N+3,2)}(\fu)$ and $\CT_{(2N+3,2)}(\fu)$ are generalized modular transformation matrices of $(2N+3,2)$ models which encodes the modular properties of torus one-point conformal blocks of $(2M+3,2)$ models with $M\geq N$.
As a result, one can use this relation to better understand the modular properties of torus one-point conformal blocks of minimal models, and also conjecture the expressions of more observables like supersymmetric partition functions on other manifolds and partition functions with surface defects insertion of $(A_1, A_{2N})$ AD theories using the refined CS theory. On the other hand, at least at the level of partition functions, the series of $(A_1, A_{2N})$ AD theories encodes the same information as the $SU(2)$ refined CS theory. Hence it might be viewed as an alternative approach of the $SU(2)$ refined CS theory.

There are still many interesting questions to be answered. One may consider generalizing this relation to $(A_{k-1}, A_{N-1})$ AD theories, and Coulomb indices may be identified with the partition function of $SU(k)$ refined CS with $\rt^k\rq^N=1$. Notice that $(A_{k-1}, A_{N-1})$ construction gives the same AD theory as $(A_{N-1}, A_{k-1})$. It is interesting to find the corresponding symmetry in refine CS theories. One can also try to generalize the relation to other AD theories, especially ones with both an irregular singularity and a regular singularity. The corresponding $M$-theory picture will have intersecting M5 branes instead of parallel M5-branes considered in this paper.

It is only an oberservation that there is a map between the vector space of torus one-point conformal blocks of minimal models and the Hilbert space of refined CS theory, and the they share the same modular property. It is then interesting to understand the underlining principle behind this map and obtain a more natural interpretation of the generalized modular transformation matrices. Notice that the characters of $(2N+3,2)$ models are identified as the Schur indices with defect insertions of $(A_1, A_{2N})$ AD theories \cite{Cordova:2016uwk,Cordova:2017mhb,Cordova:2017ohl,Neitzke:2017cxz}. It is interesting to find a similar interpretation for torus one-point conformal blocks and understand the relation between Coulomb branch indices and defected Schur indices.

Last but not least, the quadruple relation mentioned in the introduction \ref{sec:intro} predicts a map between fixed points of wild Hitchin modular space and the Hilbert space of refined CS theory, and the wild Hitchin character is equal to the refined CS partition function through the Coulomb branch index. It is also worth constructing a more precise statement of this correspondence and formulating a rigorous proof.

%The relation between the Coulomb branch indices of $(A_1, A_{2N})$ Argyres-Douglas theories on $L(k,1)\times S^1$ and the modularity of torus one-point conformal blocks of $(2N+3,2)$ minimal models is discussed in this work. These modularity data can be summarized into $\CS(\fu)$ and $\CT(\fu)$ matrices with rank greater or equal to two which satisfy
%\begin{equation}
%\begin{split}
%&\CS(\fu)^2=1,\\
%&(\CS(\fu)\CT(\fu))^3=1.
%\end{split}
%\end{equation}
%The Coulomb branch indices are proportional to the $00$ component of $\CS(\fu)^{-1}\CT(\fu)^{-k}\CS(\fu)$, which can be interpreted as a topological invariant of $L(k,1)$.
%
%The $\CS(\fu)$ and $\CT(\fu)$ matrices can also be obtained from a specialization of refined Chern-Simons theory, leading to a relation between Argyres-Douglas theories and refined Chern-Simons theory. This correspondence opens up possibilities to compute the supersymmetric partition function of Argyres-Douglas theories on other four-manifolds and add yet another player into the relations among AD theories, VOA and Hitchin system.

\section*{Acknowledgment}

The authors would especially like to thank Thomas
Creutzig and Du Pei for extensive communication and discussion. The authors would also like to thank Tomoyuki Arakawa, Chris Beem, Chi-Ming Chang, Zongbin Chen, Pavel Etingof, Davide Gaiotto, Victor Kac, Peter Koroteev, Conan Leung, Si Li, Bong Lian, Kazunobu Maruyoshi, Leonardo Rastelli, Mauricio Romo, Peng Shan, Shu-Heng Shao, Jaewon
Song, Cumrun Vafa, Dan Xie and Shing-Tung Yau for helpful discussion. The work of CK is supported by Department of Physics, Bo\u{g}azi\c{c}i University and CMSA, Harvard University. The work of WY is supported by YMSC, Tsinghua University and CMSA, Harvard University.

\appendix

\section{S-matrix}
\label{sec:Smatrixsol}

In this section explicit forms of $\CS_{(2N+3,2)}(\fu)$ are given for small $N$. More details and general solutions for arbitrary diagonal $\CT$'s are explained in \cite{Itoyama:2012re}.

Starting from $N=1$ when $\CS_{(5,2)}$ is a two by two matrix,
\begin{equation}
\CS_{(5,2)}(\fu)
=\frac{1}{1-\fu}
\left(
\begin{array}{cc}
-i\sqrt{\fu} & \sqrt{1-\fu+\fu^2} \\
\sqrt{1-\fu+\fu^2} & i\sqrt{\fu}
\end{array}\right).
\end{equation}

When $N=2$,
\begin{equation}
\CS_{(7,2)}(\fu)=
\left(
\begin{array}{ccc}
 -\frac{\fu^2}{(\fu-1)^2 \left(\fu^2+\fu+1\right)} & -\frac{\sqrt{\fu^2+\fu}\sqrt{\fu^5+1}}{\sqrt{\fu-1} \left(\fu^2-1\right)
   \sqrt{\fu^3-1}} & -\frac{\sqrt{\fu^4+1} \sqrt{\fu^5+1}}{\sqrt{\fu-1} \sqrt{\fu^2-1} \left(\fu^3-1\right)} \\
 \frac{\sqrt{\fu^2+u}  \sqrt{\fu^5+1}}{\sqrt{\fu-1} \left(\fu^2-1\right) \sqrt{\fu^3-1}} & 1+\frac{\fu}{1-\fu}+\frac{\fu^2}{(1-\fu)^2}
   & \frac{\sqrt{\fu^4+1} \sqrt{\fu^2+\fu}}{(\fu-1) \sqrt{\fu^2-1} \sqrt{\fu^3-1}} \\
 -\frac{\sqrt{\fu^4+1} \sqrt{\fu^5+1}}{\sqrt{\fu-1} \sqrt{\fu^2-1} \left(\fu^3-1\right)} & -\frac{\sqrt{\fu^4+1}
   \sqrt{\fu^2+\fu}}{(\fu-1) \sqrt{\fu^2-1} \sqrt{\fu^3-1}} & -\frac{\fu \left(\fu^2+1\right)}{(\fu-1)^2 \left(\fu^2+\fu+1\right)}
   \\
\end{array}
\right).
\end{equation}

$N=3$,
\begin{equation}
\CS_{(9,2)}(\fu)
=\left(
\begin{array}{cccc}
 U_{11} & -iU_{12} & -iU_{13} & -iU_{14} \\
 -iU_{21} & -U_{22} & iU_{23} & -iU_{24} \\
 -iU_{31} & iU_{23} & U_{33} & iU_{34} \\
 -iU_{41} & -iU_{42} & U_{43} & -U_{44}
\end{array}
\right),
\end{equation}
with
\begin{equation}
U_{ij}^2=\frac{(\xi_i^2-1)(\xi_j^2-1)\prod_{k\neq i,j}(\xi_j\xi_k-1+(\xi_j\xi_k)^{-1})}{(\xi_i-\xi_j)^2\prod_{k\neq i,j}(\xi_i-\xi_k)(\xi_j-\xi_k)},
\end{equation}
and
\begin{equation}
U^2_{ii}=1+\sum_{j\neq i} U^2_{ij}.
\end{equation}
The $\xi_i$'s are defined as,
\begin{equation}
\xi_i=(\CT_{(9,2)})_{ii}.
\end{equation}

$N=4$,
\begin{equation}
\CS_{(11,2)}(\fu)
=\left(\begin{array}{ccccc}
U_1 & -U_{12} & -U_{13} & U_{14} & U_{15} \\
-U_{12} & -U_2 & U_{23} & U_{24} & -U_{25} \\
-U_{13} & U_{23} & U_3 & -U_{34} & -U_{35} \\
U_{14} & U_{24} & -U_{34} & -U_4 & U_{45} \\
U_{15} & -U_{25} & -U_{35} & U_{45} & U_{5}
\end{array}\right),
\end{equation}
with
\begin{equation}
U_{ij}^2
= -\frac{\xi_i\xi_j(\xi_i+1+\xi^{-1}_i)(\xi_j+1+\xi_j^{-1})}{(\xi_i-\xi_j)^2}\frac{
\prod_{k\neq i,j}(1+\xi_i\xi_k)(1+\xi_j\xi_k)}
{\prod_{k\neq i,j}(\xi_i-\xi_k)(\xi_j-\xi_k)},
\end{equation}
and
\begin{equation}
U_i=1-\sum_{j\neq i}U^2_{ij}.
\end{equation}
$\xi_i$'s are the diagonal elements of $\CT_{(11,2)}$,
\begin{equation}
\xi_i=(\CT_{(11,2)})_{ii}=e^{\frac{2\pi i}{3}}\fu^{-6+\frac{9}{2}i-\frac{i^2}{2}}.
\end{equation}
The sign difference from \cite{Itoyama:2012re} in the above formula is originated from the sign difference in the $\det\CT_{(11,2)}$.

\section{Useful formulas on Macdonald polynomials}
\label{app:sec:MDpoly}

The Macdonald polyonmials depend on two parameters $\rq$ and $\rt$, where $t=q^\beta$ and $\beta\in\C^{\ast}$ is the deformation parameter. These polynomials are remarkably simple in rank one case ($SU(2)$),
\begin{equation}
P_j(x_1,x_2)=\sum_{l=0}^j x^{j-l}_1x^l_2
\prod^{l-1}_{i=0}\frac{[j-i]}{[j-i+\beta-1]}
\frac{[i+\beta]}{[i+1]},
\end{equation}
with $[x]=\frac{q^{x/2}-q^{-x/2}}{q^{1/2}-q^{-1/2}}$.

$g_i$ is the quadratic norm of the Macdonald polynomials under a natural orthogonality condition,
\begin{equation}
g_i=\prod_{m=0}^{i-1}\frac{[i-m]}{[i-m+\beta-1]}
\frac{[m+2\beta]}{[m+\beta+1]}.
\end{equation}

$P_j(\rt^\half,\rt^{-\half};\rq,\rt)$ is also called the $(\rq,\rt)$-deformed dimension of the spin-$j/2$ representation. When $t^2q^k=1$, it has the following vanishing conditions,
\begin{equation}
P_j(\rt^\half,\rt^{-\half};\rq,\rt)=0, \,\,\,\,\forall\,j>k.
\end{equation}
There is another vanishing condition which is more commonly used in the literature. For $K\in\Z^+$, $\rq=\exp(\frac{2\pi i}{K+2\beta})$ and $\rt=rq^{\beta}$,
\begin{equation}
P_j(\rt^\half,\rt^{-\half};\rq,\rt)=0, \,\,\,\,\forall\,j>K.
\end{equation}

\section{Similarity transformation in two dimensional case}
\label{app:sec:similarity}

The transformation matrix $\Omega_1$ which rotates $S_{ij}(\fu^{-2},\fu^3)$ into the upper diagonal block while keeps $T_i$ invariant is derived in this section.

Starting from the $S$ operator, Eq. \ref{eq:Srep2dim}, construct the similarity transformation matrix $\Xi$ from its eigenvectors,
\begin{equation}
\Xi=
\left(
\begin{array}{cccc}
 -\frac{(\fu+1)^3 ((\fu-1) \fu+1) \left(\fu^2+1\right)}{\fu^{7/2}} & -\frac{(\fu+1)^3 (\fu^2- \fu+1)
   \left(\fu^2+1\right)}{\fu^{7/2}} & 0 & \frac{(\fu+1)^3 (\fu^2- \fu+1) \left(\fu^2+1\right)}{\fu^{7/2}} \\
 \frac{\left(-i \fu+\sqrt{\fu}+i\right) (\fu+1)^2 \left(\fu^2+1\right)}{\fu^{5/2}} & \frac{\left(i \fu+\sqrt{\fu}-i\right)
   (\fu+1)^2 \left(\fu^2+1\right)}{\fu^{5/2}} & -\frac{(\fu+1) \left(\fu^2+1\right)}{\fu^{3/2}} & 0 \\
 -i \fu+\sqrt{\fu}+\frac{1}{\sqrt{\fu}}+\frac{i}{\fu} & i \fu+\sqrt{\fu}+\frac{1}{\sqrt{\fu}}-\frac{i}{\fu} & 1 & 0 \\
 1 & 1 & 0 & 1 \\
\end{array}
\right),
\end{equation}
and $S$ and $T$ becomes,
\begin{equation}
\begin{split}
\tilde{S}&=\Xi^{-1}\,S\,\Xi=-\frac{2i(1-\fu)}{\sqrt{\fu}}\left(
\begin{array}{cccc}
1 &0&0&0\\
0&-1&0&0\\
0&0&0&0\\
0&0&0&0
\end{array}\right),\\
\tilde{T}&=\Xi^{-1}\,T\,\Xi=
\left(
\begin{array}{cccc}
 \frac{1}{2} \left(u+i \sqrt{u}+1\right) & \frac{1}{2} \left(-u+i \sqrt{u}+1\right) & 0 & 0 \\
 \frac{1}{2} \left(-u-i \sqrt{u}+1\right) & \frac{1}{2} \left(u-i \sqrt{u}+1\right) & 0 & 0 \\
 0 & 0 & u & 0 \\
 0 & 0 & 0 & 1 \\
\end{array}
\right).
\end{split}
\end{equation}

Now use the transformation $\Pi$ to diagonalize $\tilde{T}$ while keeps the block structure of $\tilde{S}$,
\begin{equation}
\Pi=\left(
\begin{array}{cccc}
 1-\frac{2 i \sqrt{u}}{u+i \sqrt{u}-1} & -1 & 0 & 0 \\
 1 & 1 & 0 & 0 \\
 0 & 0 & 1 & 0 \\
 0 & 0 & 0 & 1 \\
\end{array}
\right),
\end{equation}
and obtain $S'$ and $T'$ in Eq.'s \ref{eq:SS2dim} and \ref{eq:TT2dim},
\begin{equation}
\begin{split}
S'&=\Pi^{-1}\,\tilde{S}\,\Pi,\\
T'&=\Pi^{-1}\,\tilde{T}\,\Pi.
\end{split}
\end{equation}

The transformation matrix $\Omega_1=\Xi\,\Pi$, and has the explicit form.
\begin{equation}
\small
\Omega_1=\left(
\begin{array}{cccc}
 -\frac{2 (\fu-1) (\fu^2+1) (\fu+1)^3 \left(\fu-i
   \sqrt{\fu}-1\right)}{\fu^{7/2}} & 0 & 0 & \frac{(\fu+1)^3 (\fu^2- \fu+1) \left(\fu^2+1\right)}{\fu^{7/2}} \\
 0 & -\frac{2 i \left(\fu-i \sqrt{\fu}-1\right) \left(\fu^5+\fu^4-\fu-1\right)}{\fu^{5/2} \sqrt{\fu^2-\fu+1}} & -\frac{(\fu+1)
   \left(\fu^2+1\right)}{\fu^{3/2}} & 0 \\
 0 & -\frac{2 i \left(\fu-i \sqrt{\fu}-1\right) \left(\fu^2-1\right)}{\fu \sqrt{\fu^2-\fu+1}} & 1 & 0 \\
 \frac{2 (\fu-1)}{\fu+i \sqrt{\fu}-1} & 0 & 0 & 1 \\
\end{array}
\right).
\end{equation}
The $(i,j)$ entries of $\Omega_1$ are non-zero only when $i=j$ or $i=3-j$, and
\begin{equation}
 (\Omega_1^{-1})_{00}(\Omega_1)_{00}=\half.
\end{equation}

\providecommand{\href}[2]{#2}

\bibliographystyle{JHEP}
\bibliography{refs}

\address{
Can Koz\c{c}az \\
Department of Physics, Bo\u{g}azi\c{c}i University, \\
Istanbul, T\"{u}rkiye\\
\email{can.kozcaz@boun.edu.tr}
}

\address{
Center for Mathematical Sciences and Applications, \\
Harvard University, \\
Cambridge, MA 02138, USA
}

\address{
Shamil Shakirov\\
Society of Fellows, Harvard University, \\
Cambridge, MA 02138, USA
}

\address{
Mathematical Sciences Research Institute, \\
Berkeley, CA 94720, USA
}

\address{
Institute for Information Transmission Problems, \\
Moscow, 127994, Russia \\
\email{shakirov.work@gmail.com}
}

\address{
Wenbin Yan \\
Yau Mathematical Science Center, Tsinghua University, \\
Haidian District, Beijing 100084, China \\
\email{wbyan@math.tsinghua.edu.cn}
}

\address{
Center for Mathematical Sciences and Applications, \\
Harvard University, \\
Cambridge, MA 02138, USA
}

\end{document}